\documentstyle[aps]{revtex}

\begin{document}
\def\pp{$pp$ }
\def\app{$\overline{p}p$ }
\def\spp{$\sigma_t(pp)$ }
\def\sapp{$\sigma_t(\overline{p}p)$ }

\draft
\preprint{}

\title{\bf A conjecture on Centauro species\thanks{\rm {To be published Proc. VI Hadron Physics - 1998, 16-21 March, Florian\'opolis, SC, Brazil.}}}

\author{M. J. Menon}

\vskip 0.3truecm

\address{Instituto de F\'{\i}sica `` Gleb Wataghin ''\\
Universidade Estadual de Campinas, Unicamp\\
13083-970 Campinas, SP, Brasil\\}

\maketitle

\begin{abstract}

It is argued that Centauro events observed in 
cosmic ray experiments
may be characteristic of only $pp$ and not $\overline{p}p$ interactions.

\end{abstract}

\vskip 0.5truecm

The failure of ``artificial production'' of Centauro events \cite{centauro}
 in $\overline{p}p$ machines \cite{machines}, led to some conjectures on the
 origin of the phenomena, such as genetic correlations, nuclear effects by 
heavy nucleus, extragalactic source, and others.
However, all these conjectures exclude one possibility: that the unusual 
events could only be associated with \pp interactions at sufficiently high 
energies and not with \app. Generically, it does not seem unnatural to think
 that an exceedingly wide production associated with high energy matter-antimatter
 annihilation ($\overline pp$) could be compensated by some kind of suppression
 effect (pionic component) in matter-matter reactions ($pp$).
In this communication we shall briefly {\it discuss this conjecture based on one
 type of physical quantity} determined from both cosmic ray and accelerator 
experiments, namely the {\it total cross section}.

Since accelerator information on $pp$ interactions are available only below 
the limiar of production \cite{centauro}, the high mean transverse momentum 
and multiplicities (characteristics of Centauro events) observed through the
 emulsion chambers, and not present at lower energies, could originate some
 additional contributions to the total \pp cross sections, that is, higher
values than general extrapolations from lower (accelerator) energies usually
 show.
Equivalently, if this extra contribution is present only in the \pp channel 
at sufficiently high energies and not in the \app one, we could expect {\it 
total \pp cross sections higher than the \app ones at the energies where
 Centauro events are observed}, namely, $\sqrt s \geq 500\ GeV$. For this 
reason we shall consider this feature as a possible signal for Centauro 
production.

Figure 1 shows experimental information presently
available on $pp$ and $\overline{p}p$ total cross sections
at the highest energy region from both accelerator and cosmic ray experiments.
 In the last case,  informations
on \spp \cite{fly} comes from proton-air cross sections. It is
 important to observe that {\it antiprotons are not expected to have any
 significant role in the bulk of these informations} and this is a basic point
 in
our argument.
Now, since what is extracted in these experiments is the proton-air cross 
section, the determination of the \pp cross section depends, in particular,
 on nuclear model 
assumptions \cite{fly,gaisser} and this has originated some puzzles and
 discrepances between different analysis, as clearly shown in Fig. 1. 
Concerning all these results, a central point is the fact that the analysis
 by Nikolaev seems
correct, has never been criticized and the same is true for the Gaisser, 
Sukhatme and Yodh (GSY) result. 
Therefore, as qualitatively shown in Fig. 1, if we assume that the Nikolaev's
 and GSY's results are the correct ones together with $pp$ data from 
accelerators (all black symbols in the figure), experimental
informations presently available do not rule out  an increase of \spp faster 
than \sapp above $\sim 500\ GeV$. We could see this as a signature of an 
additional
contribution in the $pp$ channel, not observed up to the highest energy 
accelerator
data and not present in the $\overline{p}p$ channel.

Concerning this unusual possibility of a crossing envolving \spp and \sapp,
 we recall that from the general principles of QFT, if the Froissart-Martin
 bound is saturated then
asymptoticaly we may have \cite{cross}

\begin{equation}
\Delta \sigma_t = |\sigma_t(\overline{p}p) - \sigma_t(pp)| \leq
C{\sigma_t(\overline{p}p) + \sigma_t(pp) \over logs} \sim Clogs,
\end{equation}
leading to a wide range of possible  extrapolations, including the
crossing.
Also, the Odderon hypothesis \cite{odderon} (C-odd Regge trajectory) predicts
 $\Delta \sigma \sim logs$ and the approach satisfies analyticity and unitarity.
 We conclude that general principles and some theoretical approaches, as the 
Odderon, do not rule out the possibility of an additional contribution in the
 \spp, leading
to a faster increase with the energy than \sapp.

Concerning ``model independent'' fits to experimental data based on dispersion
 relations \cite{augier}, we recall that dispersion relations are rigorous 
consequences of local QFT once one assumes analyticity, crossing, unitarity 
and, the less quoted, polynomial boundedness.
However, it has been shown by Khuri \cite{khuri} that a failure of the last 
assumption at future high energy accelerators can lead to a breakdown of QFT
and dispersion relations, the fundamental tool in ``model-independent'' analysis
 of \spp and \sapp data.

All the facts presented in this short communication do not ruled out the possibility
 that an additional contribution to \spp may be present at the highest energies 
and that it is absent in both \spp at lower energies and in the $\overline{p}p$ 
channel in all the energies.
 Our conjecture attributes this asymmetry to the unusual events observed in 
cosmic-ray hadronic interactions, the Centauro species. This could explain why
 no signals of Centauro events were observed in $\overline{p}p$ colliders.

\vskip 0.5truecm
\leftline{Acknowledgments}

I am grateful to E. H. Shibuya, C. O. Escobar and C. D. Chinellato
for valuable comments and also to A. F. Martini.

\leftline{\bf Figure caption}

\begin{figure}
\caption{Accelerator experimental data on \sapp and
\spp in the interval $13.8\ GeV \protect\leq \protect\sqrt{s} 
\protect\leq 1.8\ TeV$ and cosmic-ray results \protect\cite{fly} 
on \spp .}
\label{}
\end{figure}

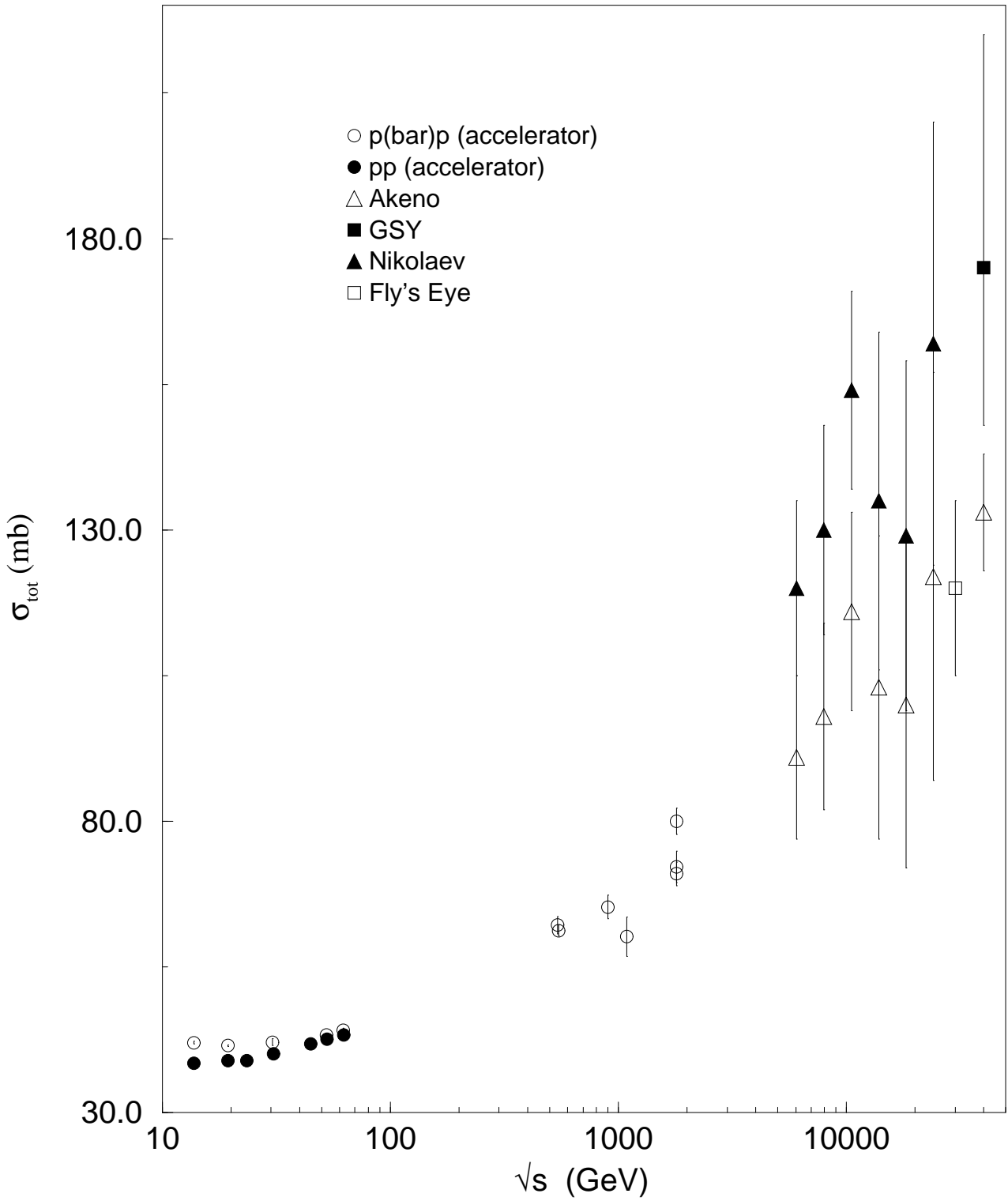
\end{document}